\documentstyle[12pt]{article}
\begin{document}
\begin{center}
{\large\bf Regge Trajectories and Renormalization Group}\\
\vskip 3mm
Vladimir A.Petrov\\
\vskip 2mm
{\it Division of Theoretical Physics, IHEP,\\
Protvino, Russia}
\vskip 5mm
\bf Abstract\\
\end{center}
\vskip 5mm

On the basis of the RG invariance we show, in particular, that the
Regge trajectory intercept  cannot be calculated as a function of the
coupling constant.
\vskip 10mm
Since the early days of the complex angular momentum approach [1] to
obtaining the high-energy asymptotic behaviour of the scattering
amplitude [2] numerous attempts were undertaken to calculate the
leading $j$-plane singularities  (poles, cuts...) and their
$t$-dependence in the course of the asymptotic summation of some
"leading" terms in the perturbative expansions (see e.g. [3]).

Last 30 years are marked by widely spread hopes that the problem can
find its solution in the framework of the study of some QCD-motivated
Bethe-Salpeter-type equations [4] the kernel of which is  a
trunkated series in the QCD coupling, $\alpha_s$.

During some quite a long time the result obtained in Ref.[4] that
the leading singularity  is of the
form $$j_0 = 1 + \frac{12}{\pi}ln 2\cdot \alpha_s$$ with $\alpha_s$
assumed to be $O(10^{-1})$, was  considered, in particular, as an
explanation of the fast
rise of the structure functions at small $x$ observed at HERA.

Eight years ago it was obtained in Refs [5] that the further
 term, proportional to $\alpha^2_s$, had an unexpetedly large and
negative coefficient which invalidated the use of this value of $j_0$
for explanation of the HERA data:

\begin{equation}
j_0 \approx 1 + \frac{12 ln 2}{\pi}\alpha_s [1 -
20\frac{\alpha_s}{\pi}].
\end{equation}

Moreover, result (1), which depends on $\alpha_s$ at an unspecified
scale, exacerbated the important problem of the scale
and
renormalization scheme dependence of $j_0$. An attempt to
minimize such a dependence with use of some "optimal" choice of
renormalization conditions was undertaken in Ref. [6].

However, one has to still bear in mind a requirement of the RG
invariance of all physical observable quantities.

This fully concerns $j_0$ as a singularity  of the physical amplitude
and therefore  all attempts to find an expression for $j_0$ that only
"slightly" violates RG invariance are of the same origin as an
attempt to use "slightly" non-Lorentz invariant expression for, say,
the mass of a particle.

From the RG invariance of the scattering amplitude and, hence, its
 singularities in the $j$-plane it follows that a generic
$j$-plane
singularity (which has generally to depend on the $4$-momentum
transfer
squared,    $t) j_0(t)$ is of the form [7]:

\begin{equation}
j_0(t) = \Phi [\frac{t}{{\mu}^2}e^{K(\alpha_s)}]
\end{equation}
where $\Phi$ is some still unknown function,
$K'(\alpha_s)=1/\beta(\alpha_s)$, $\mu^2
\frac{\partial}{\partial\mu^2}\alpha_s = \beta(\alpha_s)$, $\mu$ is
the
renormalization scale. Eq.(2) explicitly
displays the manifest RG invariance of $j_0(t)$.

Eq.(2) allows to make some interesting observations. Let us assume
t=0
with $\mu^2$ fixed and non-zero. Then 

\begin{equation}
j_0 (0) = \Phi (0).
\end{equation}
Eq.(3) means that $j_0$ does not depend on $\alpha_s$ at all! In
other words the number $j_0(0)$ cannot be calculated by any
manipulation with perturbative expansions.

On the other hand if one assumes $t=-\mu^2$ then

\begin{equation}
j_0 (t) = f (\alpha_s(t)).
\end{equation}

From Eq.(4) one can see that to obtain $j_0(0)$ from theory one has
to know the value of $\alpha_s(0)$. Straightforward (but never
proved)
confinement requirement is $\alpha_s(0)=\infty$, and this brings  us
back to the situation discussed below Eq.(3). In the literature there
exists an approach, so-called "analytical QCD", in which, on the
basis
of microcausality, a finite $\alpha_s(0)$ is obtained [8]. For
self-consistency we have to  conclude that

\begin{equation}
\Phi(0) = \Phi (e^{K[\alpha_s(0)]}),
\end{equation}
i.e. $K[\alpha_s(0)] = -\infty$ and $\beta$-function has to vanish at
$\alpha = \alpha_s(0)$.

The facts discussed above are not very encouraging for those who try
to obtain  $j_0(t)$ from various B-S-like equations for the
scattering  amplitude. At any rate it is mandatory  to work with
RG-invariant BS-kernels. The latter is an extremly difficult task and
this, as one can suppose, underlies the discussion of the
diffractive scattering in terms of the $N=4$ supersymmetric version
of QCD, where, as is well known, $\beta=0$, and above mentioned
problems seem to be lifted [9].  Inviting supersymmetry is by no
means
motivated by the available data which
in no way indicate even a tiny  trace of supersymmetry. In
principle one could expect such traces in high-energy collisions in
the events with very large energy-momentum transfers what is
definitely not the case for diffractive processes. The same can be
said about recent attempts  to use the string theory. 

Does it mean that the diffractive processes are hopeless in the sense
of their
study in the framework of QCD? Not at all, but this study requires
seriuos non-perturbative insights. Let us take, for instance, the
forward scattering amplitude $T(s,0)$. Due to RG invariance

\begin{equation}
T(s,0) = F(\alpha_s(s))
\end{equation}

On the other hand the existing data seem to indicate that the
forward scattering amplitude behaves
asymptotically  as

\begin{equation}
T(s,0) \sim i \>const \>s\> log^2 s + ...
\end{equation}

Such a behaviour corresponds to the asymptotics, obtained in some
model by W. Heisenberg [10], and later was proved to be a general
upper bound
[11]. Asymptotic behaviour of the type of (7) was later obtained  in
a
field-theoretic  model for the
eikonal [12]. 
In terms of the running coupling $\alpha_s(s)$ it could look as an
awful "antiperturbative" dependence:

\begin{equation}
T(s,0) \sim \>i\> const\> \frac{1}{\alpha^2_s(s)}e^{\frac{1}{\beta_0
\alpha_s(s)}}.
\end{equation}
which clearly shows that  in this case one can hardly obtain such a
behaviour from an asymptotic series in $\alpha_s(s)$. Rather it
could be
 obtained from some effective field configuration which gives a
 leading extremum of the action in the path-integral  representation
 for
the scattering amplitude [13]:

$$
T(s,t) \sim \int d A_{\mu}dqd\bar q e^{i\frac{S[A,q,\bar
q]}{\alpha_s}}
\tilde H(p_1)\tilde H(p_2)\tilde H(p_3)H(0),
$$
where $H(x) = \int \frac{d4p}{(2\pi)^4} e^{-ipx}\tilde H(p)$ is the
Heisenberg source field composed of the elementary fields $A,q,\tilde
q$ and corresponding to hadron fields.

This possibility is in agreement with the fact that at higher
energies the effective interaction region grows and collective,
quasiclassical field configurations seem to be more relevant than any
description in terms of elementary excitations (gluons, quarks).

Eq. (8) looks as a paradox, because it seems quite natural that
$T\rightarrow 0$ if we take the free-field limit $\alpha_s
\rightarrow
0$. Two options can take place. The first is to agree that, contrary
to
(7), $T(s,0) \rightarrow 0$, because $\alpha_s \rightarrow 0$
corresponds to $s \rightarrow \infty$. The second is to admit  that
the
limit $\alpha_s \rightarrow 0$ does not commute with the path
integration.

\vskip 5mm
{\bf Acknowledgements}

I thank A.I. Alekseev, A.K.Likhoded, A.A.Logunov and N.E. Tyurin for
useful conversations on various aspects of the subject of this work.

\newpage

{\bf References}

1. T. Regge. Nuovo Cimento, \underline{8}, 671 (1959)

2. G.F. Chew and S.C. Frautschi. Phys.Rev.Lett. \underline{7} (1961)
394

3. B.W. Lee and R.F. Sawyer. Phys.Rev. \underline{1}27 (1962) 2266;

B.A. Arbuzov, A.A. Logunov, A.N. Tavkhelidze Phys.Lett.
\underline{2}(1962) 150

4. L.N. Lipatov. Sov.J.Nucl.Phys. \underline{23} (1976) 338;

E.A. Kuraev, L.N. Lipatov and V.S. Fadin. Sov.Phys. JETP

\underline{44} (1976) 443.

Ya.Ya. Balitzkij and L.N. Lipatov. Sov.J.Nucl.Phys. \underline{28}
(1978) 822.

5.  V.S. Fadin and   L.N. Lipatov. Phys.Lett. B429 (1998) 127;

G. Camici and M.Ciafaloni. Phys.Lett. B430	(1998) 349

6. S.J. Brodsky, V.S. Fadin, V.T. Kim, L.N. Lipatov and 

G.V.
Pivovarov. JETP Lett. 70 (1999)155

7. N.N. Bogolyubov and D.V. Shirkov. Introduction to the theory of

quantized fields. Wiley, New York, 1980.

G.B. West. Phys.Rev. D27(1983) 1402.

8. D.V. Shirkov and I.L. Solovtsov.  Phys.Rev.Lett. 79 (1997) 1209. 

The stability of finite $\alpha_s(0)$ with respect to all higher loop
contributions 

was shown  by  A.I. Alekseev. Few Body Syst. 32 (2003)
193.

9. A.V. Kotikov and L.N. Lipatov. Nucl. Phys. B661 (2003) 19

10. W. Heisenberg. Z.Phys. 133 (1952) 65

11. M. Froissart. Phys.Rev. 123 (1961) 1053

A. Martin. Nuovo Cim. 42 (1966) 901

12. H. Cheng and T.T. Wu. Phys.Rev. 182 (1969) 1852.

13. Some steps towards this goal can be found in the following
papers:

H. Moreno and H.M. Fried. Phys.Rev.  D12 (1975) 2031;

P. Suranyi. Phys.Rev. D17 (1978) 1160.
\end{document}